\documentclass[%
reprint,
superscriptaddress,
twocolumn,
notitlepage,
nofootinbib,
amsmath,amssymb,
aps,
pra,
]{revtex4-1}

\usepackage{graphicx}% Include figure files
\usepackage{dcolumn}% Align table columns on decimal point
\usepackage{bm}% bold math
\usepackage{tablefootnote}

\usepackage{amsthm}
\usepackage{amsmath}
\usepackage{amssymb}
\usepackage{graphicx}
\usepackage{url}
\usepackage{float}
\usepackage{bm}
\usepackage{ifthen}
\usepackage[usenames,dvipsnames]{color}
\usepackage{mathrsfs}
\usepackage[colorlinks=true,citecolor=blue,urlcolor=black]{hyperref}
\usepackage{float}
\usepackage{subfigure}
\usepackage{enumitem}
\usepackage{color}
\usepackage{setspace}
\usepackage{braket}
\usepackage{comment}
\usepackage{lipsum}
\newcommand{\be}{\begin{equation}}
\newcommand{\ee}{\end{equation}}

\usepackage{mathtools}

\begin{document}
	
	%\tableofcontents
	
	\title{Simple Method for Asymmetric Twin-Field Quantum Key Distribution}
	\author{Wenyuan Wang}
	\affiliation{Centre for Quantum Information and Quantum Control (CQIQC), Dept. of Electrical \& Computer Engineering and Dept. of Physics, University of Toronto, Toronto,  Ontario, M5S 3G4, Canada}
	
	\author{Hoi-Kwong Lo}
	\affiliation{Centre for Quantum Information and Quantum Control (CQIQC), Dept. of Electrical \& Computer Engineering and Dept. of Physics, University of Toronto, Toronto,  Ontario, M5S 3G4, Canada}
	
	\begin{abstract}
		Twin-Field quantum key distribution (TF-QKD) can beat the linear bound of repeaterless QKD systems. After the proposal of the original protocol, multiple papers have extended the protocol to prove its security. However, these works are limited to the case where the two channels have equal amount of loss (i.e. are symmetric). In a practical network setting, it is very likely that the channels are asymmetric due to e.g. geographical locations. In this paper we extend the "simple TF-QKD" protocol to the scenario with asymmetric channels. We show that by simply adjusting the two signal states of the two users (and not the decoy states) they can effectively compensate for channel asymmetry and consistently obtain an order of magnitude higher key rate than previous symmetric protocol. It also can provide 2-3 times higher key rate than the strategy of deliberately adding fibre to the shorter channel until channels have equal loss (and is more convenient as users only need to optimize their laser intensities and do not need to physically modify the channels). We also perform simulation for a practical case with three decoy states and finite data size, and show that our method works well and has a clear advantage over prior art methods with realistic parameters.

	\end{abstract}
	
	\date{\today}
	\maketitle

%\section{Introduction}

\section{Background}

Quantum key distribution (QKD) is proven to provide information-theoretic security to two communicating parties. Without efficient quantum repeaters, however, QKD is limited in the maximum distance over which it can generate secure keys. The linear (also named PLOB) bound is a theoretical upper bound for maximum key rate-distance relation for repeaterless QKD. Interestingly, the Twin-Field (TF) QKD protocol proposed in 2018 \cite{TFQKD} uses a clever technique to surpass the linear bound: it uses a setup where two parties, Alice and Bob, communicate with an untrusted third party, Charles. Instead of using two-photon interference like in measurement-device-independent (MDI) QKD \cite{mdiqkd}, TF-QKD makes use of single-photon interference to generate keys, and on average only one photon passes through either Alice's or Bob's channel - which allows to key rate to scale with transmittance over only half the distance between Alice and Bob. Not only does TF-QKD surpass the repeaterless bound, it also provides security against attacks on measurement devices \cite{devices_review} similar to MDI-QKD. Because of these advantages, TF-QKD has attracted much attention worldwide since its proposal. Since a rigorous security proof is not provided in the original proposal, several papers have improved the protocol and provided security proof \cite{TFQKD01,TFQKD02,TFQKD03,simpleTFQKD,TFQKD04}. Also, recently there have been multiple reports of TF-QKD demonstrated experimentally \cite{TFexperiment01,TFexperiment02,TFexperiment03,TFexperiment04}.

However, all the above security proofs and experimental demonstrations only consider the symmetric case where Alice's and Bob's channels have the same amount of loss. In reality, though, in a network setting, due to e.g. geographical locations, or Alice and Bob being situated on moving free-space platforms (such as ships or satellites), it is very likely that Alice's and Bob's channels are not symmetric. In the future, if a quantum network is build around the protocol - e.g. a star-shaped network where numerous users (senders) are connected to one central node with measurement devices, asymmetry will be an even more severe problem since it is difficult to maintain the same channel loss for all users (and users might join/leave a network at arbitrary locations). If channels are asymmetric, for prior art protocols, users would either have to suffer from much higher quantum-bit-error-rate (QBER) and hence lower key rate, or would have to deliberately add fibre to the shorter channel to compensate for channel asymmetry, which is inconvenient (since it requires physically modifying the channels) and also provides sub-optimal key rate.

Similar limitation to symmetric channels have been observed in MDI-QKD. In Ref. \cite{mdi7intensity}, we have proposed a method to overcome this limitation, by allowing Alice and Bob to adjust their intensities (and use different optimization strategies for two decoupled bases) to compensate for channel loss, without having to physically adjust the channels. The method has also been successfully experimentally verified for asymmetric MDI-QKD in Ref. \cite{asymmetric_experiment}. 

In this work, we will apply our method to TF-QKD and show that it is possible to obtain good key rate through asymmetric channels by adjusting Alice's and Bob's intensities - in fact, we will show that, Alice and Bob only need to adjust their signal intensities to obtain optimal performance. We show that the security of the protocol is not affected, and that an order of magnitude higher (than symmetric protocol) or 2-3 times higher (than adding fibre) key rate can be achieved with the new method. Furthermore, we show with numerical simulation results that our method works well for both finite-decoy and finite-data case with practical parameters, making it a convenient and powerful method to improve the performance of TF-QKD through asymmetric channels in reality.

The idea of asymmetric TF-QKD protocol has also been discussed in a recent paper \cite{tf01}. Our work is different from Ref. \cite{tf01} in several aspects. First, Ref. \cite{tf01} starts with a different protocol---``sending or not sending protocol". Second, Ref. \cite{tf01} is mostly numerical. In contrast, we start with the ``simple TF-QKD" protocol in \cite{simpleTFQKD}  and consider its asymmetric version. We include both analytical and numerical reasoning. We also provide a detailed discussion about the physics behind the security of our asymmetric protocol.

\section{Protocol}

Here we consider a similar Twin-Field (TF) QKD setup as in Ref. \cite{simpleTFQKD} ``Protocol 3". Alice and Bob choose two bases X and Z randomly. When X basis is chosen, Alice (Bob) sends states $\ket{\alpha}_a$ ($\ket{\alpha}_b$) for bit $b_A=0$ ($b_B=0$) or states $\ket{-\alpha}_a$ ($\ket{-\alpha}_b$) for bit $b_A=1$ ($b_B=1$). When Z basis is chosen, Alice and Bob send phase-randomized coherent states $\rho_{a,\beta_A}$ ($\rho_{b,\beta_B}$), where the decoy state intensities are $\{\beta_A,\beta_B\}$. Note that here Alice and Bob have a common phase reference for X basis signals. After the signals are sent to Charles, the detector events are denoted as $k_c,k_d$ (0 denotes no click, and 1 denotes a click).

\begin{figure}[h]
	\includegraphics[scale=0.3]{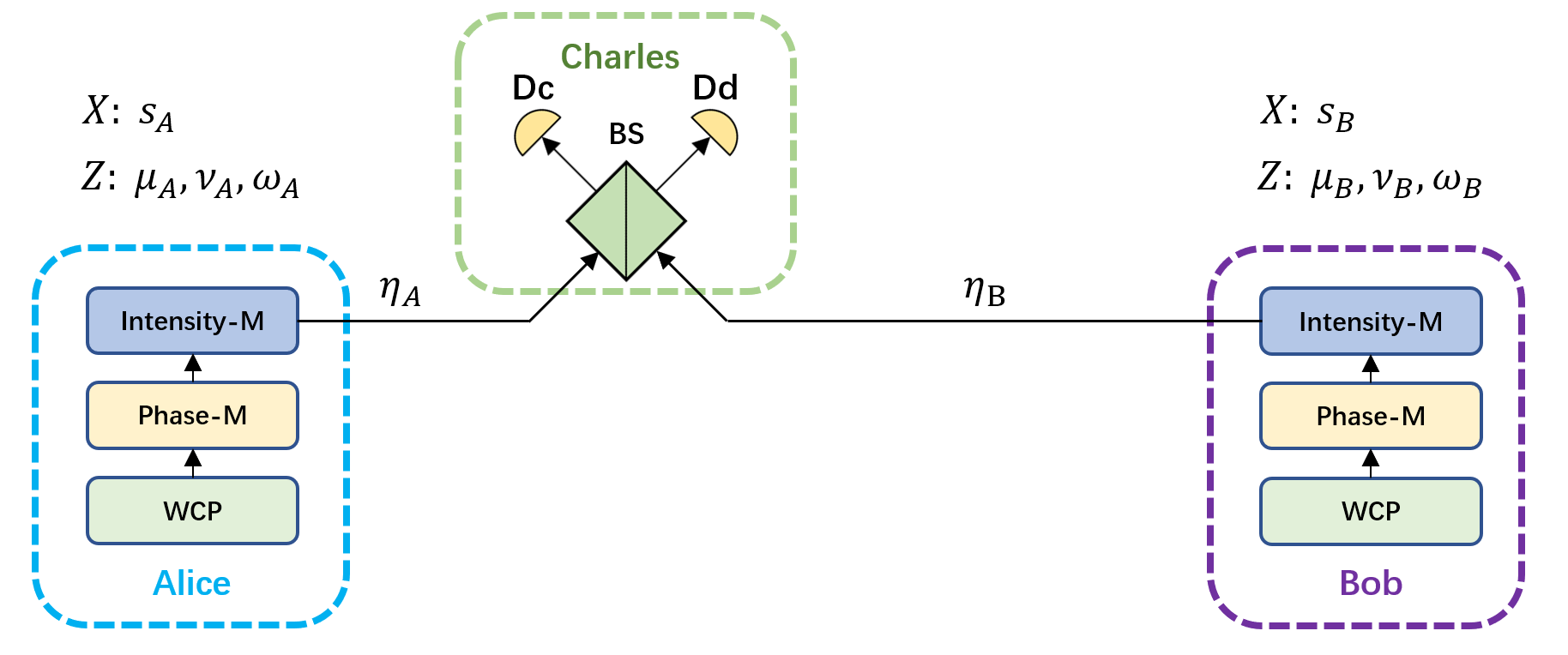}
	\caption{An example setup for a Twin-Field QKD system. Alice and Bob send signals in X and Z bases randomly. In X basis, Alice and Bob send coherent states with amplitudes $\alpha_A,\alpha_B$ (with intensities $s_A = \alpha_A^2$, $s_B = \alpha_B^2$ - in the asymmetric case, we allow $s_A$ to be different from $s_B$), phase-modulated by $\{0,\pi\}$ depending on the encoded bit. In Z basis, Alice and Bob send signals in phase-randomized coherent states, with intensities chosen from $\{\mu_A,\nu_A,\omega_A\}$ and $\{\mu_A,\nu_A,\omega_A\}$, respectively. Charles performs a swap test on the incoming signals and reports the click events in his two detectors $D_c,D_d$ (denoted by $k_c,k_d$). By choosing different intensities between Alice and Bob (and using the decoupled X and Z bases), the protocol can have high key rate even if Alice and Bob's channels have different transmittances, $\eta_A,\eta_B$.}
	\label{fig:setup}
\end{figure}

The papers \cite{TFQKD,simpleTFQKD} consider only the case where the channels between Alice (Bob) and Charles have equal transmittances. In reality, it is possible that the channels might have different levels of loss, due to e.g. geographical locations or moving platforms. Here we are interested in three questions for TF-QKD with asymmetric channels:

\noindent 1. Does channel asymmetry affect security?

\noindent 2. How does channel asymmetry affect the quantum bit error rate (QBER) and hence key rate?\footnote{In the Supplementary Materials of Ref. \cite{asymmetric_experiment}, we and our collaborators presented a preliminary study on this point, and showed that asymmetry decreases single-photon interference visibility - which will in turn increase observable QBER for TF-QKD.}

\noindent 3. Can we improve the performance of the protocol under channel asymmetry?\\

We will use our method from Ref. \cite{mdi7intensity} and apply it to Protocol 3 in Ref. \cite{simpleTFQKD}, to make an asymmetric TF-QKD protocol that works well even when channels are highly asymmetric. Similar to MDI-QKD, the protocol in \cite{simpleTFQKD} has decoupled X and Z bases. Here we allow Alice and Bob to have different intensities in the X and the Z bases respectively, such that in X basis Alice (Bob) now send states $\ket{\alpha_A}_a$ ($\ket{\alpha_B}_b$) for bit $b_A=0$ ($b_B=0$) or states $\ket{-\alpha_A}_a$ ($\ket{-\alpha_B}_b$) for bit $b_A=1$ ($b_B=1$). We can denote the signal intensities as $s_A = \alpha_A^2$, $s_B = \alpha_B^2$. In the Z basis, the amplitudes for the phase-randomized coherent states, $\{\beta_A,\beta_B\}$, can be different for Alice and Bob too (we can denote the intensities as $\{\beta_A^2,\beta_B^2\}$, and for the three-decoy case, the sets of intensities can be specifically written as $\{\mu_A,\nu_A,\omega_A\}$ and $\{\mu_A,\nu_A,\omega_A\}$). An example setup can be found in Fig. \ref{fig:setup}.

{\color{black}
We will answer the above three questions by showing in the following text three main pieces of results:\\

(1) Neither asymmetric channels nor asymmetric intensities between Alice and Bob affect security. 

(2) The X basis (signal state) QBER will increase with channel asymmetry, and greatly reduce the key rate of TF-QKD if no compensation is performed - on the other hand, the Z basis gain (as well as the upper bound to the yield and phase-error rate derived from the observable data in the Z basis) is little affected by channel asymmetry.

(3) We can use different intensities between Alice and Bob to compensate for channel asymmetry and get good key rate - in fact, using only different signal states between Alice and Bob (and keeping all decoy states and probabilities identical for Alice and Bob) can already effectively compensate for channel asymmetry and allow good key rate for asymmetric TF-QKD.
}

\section{Security}

In this section we will show that neither asymmetric channels nor asymmetric intensities between Alice and Bob affect security. Following the discussion in \cite{simpleTFQKD}, the key is generated from events in the X basis, and the secure key rate is bounded using the bit-error rate and the phase-error rate. The X basis bit-error rate is directly obtained as an observable, hence the key part of the security proof lies in the estimation of X basis phase-error rate (equivalent to the Z basis bit-error rate) based on the Z basis observables - which, since Z basis signals are phase-randomized, is not directly obtainable.

In the security proof in Ref. \cite{simpleTFQKD}, the phase-error rate is obtained by upper-bounding the phase-error rate using the observed \textit{yields} of given photon numbers $\{m,n\}$ (which can be estimated using decoy-state analysis, based on observed count rates, i.e. the gains, in the Z basis). 

The key message we'd like to point out is that, this entire estimation process of the phase error rate \textit{does not} rely on the fact that Alice and Bob use the same amplitude $\alpha$ for their signal states, or that the channels have the same transmittance. Therefore, here we will follow the proof in Ref. \cite{simpleTFQKD} step-by-step, but with asymmetric intensities and channel transmittances, to show that the security proof can be easily - in fact trivially - extended to the asymmetric case.

We can start by imagining a virtual scenario where Alice (Bob) prepares entangled states between a local qubit A (B) and a signal a (b) to be sent to Charles. After Charles performs a measurement on the signals, the X basis phase-error rate (or Z basis bit-error rate) can be obtained by Alice (Bob) measuring their local qubits in the Z basis. The initial states can be written as:

\begin{equation}
\begin{aligned}
\ket{\psi_X^A}_{Aa} &= {1\over \sqrt{2}}(\ket{+}_A\ket{\alpha_A}_a + \ket{-}_A\ket{-\alpha_A}_a)\\
\ket{\psi_X^B}_{Bb} &= {1\over \sqrt{2}}(\ket{+}_B\ket{\alpha_B}_b + \ket{-}_B\ket{-\alpha_B}_b)
\end{aligned}
\end{equation}

\noindent here $\ket{\pm}={1\over \sqrt{2}}(\ket{0}\pm\ket{1})$ are the X basis states. Here in the asymmetric case, we allow $s_A = \alpha_A^2$ to be different from $s_B = \alpha_B^2$. Now, the process of signals a and b going through their respective channels and Charles making a measurement can be represented by a Kraus operator $\hat{M}^{ab}_{k_ck_d}$, where $k_c,k_d$ are Charles' detector events. Note that, this operator $\hat{M}^{ab}_{k_ck_d}$ includes all information of the channels and detectors and is a general representation of their joint effects, and, importantly, it \textit{does not} require that the channels are symmetric at all. After signals pass through the channels and Charles makes a measurement, the state becomes:

\begin{equation}
\begin{aligned}
\ket{\chi_{k_ck_d}}_{Aa'Bb'}={{\hat{M}^{ab}_{k_ck_d}\ket{\psi_X^A}_{Aa}\ket{\psi_X^B}_{Bb}}\over{\sqrt{p_{XX}(k_c,k_d)}}}
\end{aligned}
\end{equation}

\noindent here $p_{XX}(k_c,k_d)$ is the X basis Gain for detection events $k_c,k_d$ (which can be $0,1$ or $1,0$ for a detection event to be considered successful). By measuring their local qubits in the Z basis, Alice and Bob can obtain the Z basis bit-error rate $e_{ZZ,k_ck_d}$ (i.e. the X basis phase-error rate):

\begin{equation}
\begin{aligned}
e_{ZZ,k_ck_d}=\sum_{j=0,1}||{}_{AB}\bra{jj}\ket{\chi_{k_ck_d}}_{Aa'Bb'}||^2
\end{aligned}
\end{equation}

Now, the key observation in the proof of \cite{simpleTFQKD} is that, Alice and Bob making a measurement on the local qubits A and B after sending signals a and b and Charles making a measurement should be equivalent to the time-reversed scenario where Alice and Bob first make local Z basis measurements on states $\ket{\psi_X^A}_{Aa},\ket{\psi_X^B}_{Bb}$, and then send the signal systems a and b to Charles. After Alice and Bob make the local measurements, the states become 

\begin{equation}
\begin{aligned}
{}_A\bra{0}\ket{\psi_X}_{Aa}&=\ket{C_0^A}_a\\
{}_A\bra{1}\ket{\psi_X}_{Aa}&=\ket{C_1^A}_a\\
{}_B\bra{0}\ket{\psi_X}_{Bb}&=\ket{C_0^B}_b\\
{}_B\bra{1}\ket{\psi_X}_{Bb}&=\ket{C_1^B}_b\\
\end{aligned}
\end{equation}

\noindent which are cat states:

\begin{equation}
\begin{aligned}
\ket{C_0^A}_a &= e^{-{{\alpha^2_A}\over{2}}} \sum_{n=0}^{\infty} {{\alpha^{2n}_A}\over{\sqrt{2n}}}\ket{2n}_a= \sum_{n=0}^{\infty}c_n^{A,(0)}\ket{n}_a\\
\ket{C_1^A}_a &= e^{-{{\alpha^2_A}\over{2}}} \sum_{n=0}^{\infty} {{\alpha^{2n+1}_A}\over{\sqrt{2n+1}}}\ket{2n+1}_a= \sum_{n=0}^{\infty}c_n^{A,(1)}\ket{n}_a\\
\ket{C_0^B}_b &= e^{-{{\alpha^2_B}\over{2}}} \sum_{n=0}^{\infty} {{\alpha^{2n}_B}\over{\sqrt{2n}}}\ket{2n}_b= \sum_{n=0}^{\infty}c_n^{B,(0)}\ket{n}_b\\
\ket{C_1^B}_b &= e^{-{{\alpha^2_B}\over{2}}} \sum_{n=0}^{\infty} {{\alpha^{2n+1}_B}\over{\sqrt{2n+1}}}\ket{2n+1}_b= \sum_{n=0}^{\infty}c_n^{B,(1)}\ket{n}_b\\
\end{aligned}
\end{equation}

\noindent here the even (odd) cat states only contain nonzero amplitudes for even (odd) photon numbers. Nonetheless we can still write the amplitudes as $c_n^{A,(0)}, c_n^{B,(0)}$ ($c_n^{A,(1)}, c_n^{B,(1)}$) for all photon number states, where the coefficients are zero for odd (even) photon number states in an even (odd) cat state. 

Note that here in the asymmetric case, Alice and Bob's cat states are not the same, because they use different signal intensities (hence different amplitudes $\alpha_A,\alpha_B$), but as we will show below, this does not affect the estimation of the upper bound for the phase error rate.

For Alice and Bob's local Z basis measurement results $i,j=0,1$ and for detection events $k_c,k_d$:

\begin{equation}
\begin{aligned}
&p_{XX}(k_c,k_d)||\prescript{}{AB}{\bra{ij}}\ket{\chi_{k_ck_d}}_{Aa'B'b}||^2\\
&=||\hat{M}^{ab}_{k_ck_d}\prescript{}{A}{\bra{i}}\ket{\psi_X^A}_{Aa}\prescript{}{B}{\bra{j}}\ket{\psi_X^B}_{Bb}||^2\\
&=||\hat{M}^{ab}_{k_ck_d}\ket{C_i^A}_{a}\ket{C_j^B}_{b}||^2		
\end{aligned}
\end{equation}

\noindent which means that, the probabilities for local Z basis measurement results $i,j=0,1$ (which determine the phase-error rate) can be acquired by observing the gain if Alice and Bob sent cat states. However, Alice and Bob are not really sending cat states - when Z basis is chosen, they are sending phase-randomized coherent states. Using decoy-state analysis, what Alice and Bob acquire are the yields for phase-randomized photon number states, $p_{ZZ}(k_c,k_d|n_A,n_B)$. The yields for photon number states are linked to Eq. (6) using the Cauchy-Schwarz inequality that upper-bounds the gains for cat states (and subsequently the phase-error rate):

\begin{equation}
\begin{aligned}
&||\hat{M}^{ab}_{k_ck_d}\ket{C_i^A}_{a}\ket{C_j^B}_{b}||^2\\
=&\prescript{}{a}{\bra{C_i^A}}\prescript{}{b}{\bra{C_j^B}}\hat{M}^{ab\dagger}_{k_ck_d}\hat{M}^{ab}_{k_ck_d}\ket{C_i^A}_{a}\ket{C_j^B}_{b}\\
=&\sum_{m_A,m_B,n_A,n_B=0}^{\infty}c_{m_A}^{A,(i)}c_{m_B}^{B,(j)}c_{n_A}^{A,(i)}c_{n_B}^{B,(j)} \\
& \times\prescript{}{a}{\bra{m_A}}\prescript{}{b}{\bra{m_B}}\hat{M}^{ab\dagger}_{k_ck_d}\hat{M}^{ab}_{k_ck_d}\ket{n_A}_{a}\ket{n_B}_{b}\\
\leq &\sum_{m_A,m_B,n_A,n_B=0}^{\infty}c_{m_A}^{A,(i)}c_{m_B}^{B,(j)}c_{n_A}^{A,(i)}c_{n_B}^{B,(j)} \\
& \times||\hat{M}^{ab}_{k_ck_d}\ket{m_A}_{a}\ket{m_B}_{b}||\times||\hat{M}^{ab}_{k_ck_d}\ket{n_A}_{a}\ket{n_B}_{b}||\\
=&\left[\sum_{n_A,n_B=0}^{\infty}c_{n_A}^{A,(i)}c_{n_B}^{B,(j)} \sqrt{p_{ZZ}(k_c,k_d|n_A,n_B)}\right]^2\\
\end{aligned}
\end{equation}

This means that, the phase-error rate can be upper-bounded by the yields for photon number states $p_{ZZ}(k_c,k_d|n_A,n_B)$:

\begin{equation}
\begin{aligned}
&p_{XX}(k_c,k_d)e_{ZZ}(k_c,k_d)\\
=&p_{XX}(k_c,k_d)\sum_{i=0,1}||\prescript{}{AB}{\bra{ii}}\ket{\chi_{k_ck_d}}_{Aa'B'b}||^2\\
\leq& \sum_{i=0,1}\left[\sum_{n_A,n_B=0}^{\infty}c_{n_A}^{A,(i)}c_{n_B}^{B,(i)} \sqrt{p_{ZZ}(k_c,k_d|n_A,n_B)}\right]^2\\
\end{aligned}
\end{equation}

\noindent this phase error rate, combined with the bit error rate in the X basis, can be used to perform privacy amplification on the error-corrected raw keys and obtain the secure key.

The key point is that, the above proof that upper bounds the phase error rate does not require the fact that $\alpha_A=\alpha_B$ at all. The different signal intensities will cause Alice and Bob to have different cat states, but these states are independently used to obtain inner product with $\prescript{}{A}{\bra{i}}$ and $\prescript{}{B}{\bra{j}}$ respectively. With the Cauchy-Schwarz inequality, the joint cat states are reduced to a mixture of photon number states, and there are no cross-terms between the two cat states. 

This means that, using asymmetric intensities between Alice and Bob will not affect the estimation of phase error rate. Moreover, as we described in Eq. (2), $\hat{M}^{ab}_{k_ck_d}$ is a general representation of the channels and detection, and does not require that $\eta_A=\eta_B$ either, i.e. asymmetric channels do not affect the security proof either.

Additionally, the decoy intensities $\{\beta_A^2,\beta_B^2\}$ might be different for Alice and Bob too, but these states are only used to estimate the yields of photon number states $p_{ZZ}(k_c,k_d|n_A,n_B)$ using decoy-state analysis, which is exactly the same process as in MDI-QKD. As long as Eve cannot distinguish pulses from different intensity settings, this decoy-state analysis is secure, even in the asymmetric setting - since the sending of a given photon number $n$ given the Poisson distribution $P(n|\mu)=e^{-\mu}{{\mu^n}\over{n!}}$ is a Markov process, i.e. memoryless process, Eve has no way of telling which intensity setting the photon number state came from, therefore using asymmetric intensities does not affect the estimation of yields for photon number states $p_{ZZ}(k_c,k_d|n_A,n_B)$.

Therefore, overall, we conclude that neither asymmetric channel losses, nor asymmetric intensities Alice and Bob use (for signal states or decoy states), will affect the security of the protocol. Asymmetry will only affect the performance of the protocol (which will be the subject of discussion in the next section) - asymmetric channels will result in higher QBER and subsequently lower key rate, and asymmetric intensities can compensate for channel asymmetry and enable high key rate for the protocol even when channels are highly asymmetric.

\section{Performance}

In this section we will discuss how channel asymmetry, and asymmetric intensities, can affect the performance of TF-QKD.

\subsection{Channel Model}

We will first discuss the channel model in the asymmetric case. Again, we extend the expressions in the Appendix of Ref. \cite{simpleTFQKD}, and consider asymmetric intensities and channel transmittances.

To obtain the secure key rate, three sets of observables are needed: the X basis gain $p_{XX}(k_c,k_d)$, the X basis bit-error rate $e_{XX}(k_c,k_d)$, and the Z basis gain $p_{ZZ}(k_c,k_d|\beta_A,\beta_B)$ (for all combinations of $\{\beta_A,\beta_B\}$). 

Now, let us suppose Alice and Bob send signals with intensities $s_A,s_B$, and channels between Alice/Bob and Charles have transmittances $\eta_A,\eta_B$. For simplicity we can write:

\begin{equation}
\begin{aligned}
\gamma_A &= s_A \eta_A\\
\gamma_B &= s_B \eta_B\\
\end{aligned}
\end{equation}

\noindent for signal states, and

\begin{equation}
\begin{aligned}
\gamma'_A &= \mu^i_A \eta_A\\
\gamma'_B &= \mu^j_B \eta_B\\
\end{aligned}
\end{equation}

\noindent for decoy states, where $\mu_A^i$ and $\mu_B^j$ are selected from the set of decoy intensities.

The other imperfections in the channel include the dark count rate $p_d$, the polarization misalignment between Alice and Bob $\theta$, and the phase mismatch $\phi$ between Alice and Bob. If we first do not consider dark counts and phase mismatch, the intensities arriving at the detectors C and D at Charles can be written as (similar to the discussions in Ref. \cite{mdipractical}):

\begin{equation}
\begin{aligned}
D_c &= {{1}\over{2}}\left( \gamma_A + \gamma_B - 2\sqrt{\gamma_A\gamma_B} cos\theta \right)\\
D_d &= {{1}\over{2}}\left( \gamma_A + \gamma_B + 2\sqrt{\gamma_A\gamma_B} cos\theta \right)
\end{aligned}
\end{equation}

\noindent the probability that one detector clicks and the other doesn't (e.g. C clicks and D doesn't) can be written as

\begin{equation}
\begin{aligned}
&(1-e^{-D_c})e^{-D_d}\\
=&e^{-D_d} - e^{-(D_c+D_d)}\\
=&e^{-{{1}\over{2}}\left[ \gamma_A + \gamma_B + 2\sqrt{\gamma_A\gamma_B} cos\theta \right]} - e^{-(\gamma_A+\gamma_B)}\\
\end{aligned}
\end{equation}

Including the phase mismatch and dark counts, we can write the X basis gain and QBER in a similar form as Ref. \cite{simpleTFQKD}:

\begin{equation}
\begin{aligned}
&p_{XX}(k_d,k_d)\\
=&{{1}\over{2}}(1-p_d) \left( e^{-\sqrt{\gamma_A\gamma_B} cos\phi cos\theta} + e^{\sqrt{\gamma_A\gamma_B} cos\phi cos\theta} \right)\\
\times& e^{-{{1}\over{2}}(\gamma_A+\gamma_B)} -(1-p_d)^2 e^{-(\gamma_A + \gamma_B)}\\
\end{aligned}
\end{equation}

\begin{equation}
\begin{aligned}
&e_{XX}(k_d,k_d)\\
=&{{ e^{-\sqrt{\gamma_A\gamma_B} cos\phi cos\theta}-(1-p_d)e^{-{{1}\over{2}}(\gamma_A+\gamma_B)}}  \over  {e^{-\sqrt{\gamma_A\gamma_B} cos\phi cos\theta} + e^{\sqrt{\gamma_A\gamma_B} cos\phi cos\theta}-2(1-p_d)e^{-{{1}\over{2}}(\gamma_A+\gamma_B)}}}\\
\end{aligned}
\end{equation}

\noindent and the Z basis gain is the integral over all possible (random) relative phases:

\begin{equation}
\begin{aligned}
&p_{ZZ}(k_d,k_d|\beta_A,\beta_B)\\
=&(1-p_d)\left[e^{-{{1}\over{2}}(\gamma'_A+\gamma'_B)}I_0(\sqrt{\gamma'_A\gamma'_B}cos\theta) - e^{-(\gamma'_A+\gamma'_B)}\right]\\
+&p_d(1-p_d)e^{-(\gamma'_A+\gamma'_B)}\\
\end{aligned}
\end{equation}

The Z basis gain can be used in decoy-state analysis to obtain $m,n$ photon yields $p_{ZZ}(k_c,k_d|n_A,n_B)$. Here for simplicity we first consider the infinite-decoy case, where $p_{ZZ}(k_c,k_d|n_A,n_B)$ can be assumed to be perfectly known {\color{black} (similar to Eq. (35) in Ref. \cite{simpleTFQKD} but with asymmetric channel transmittances):

\begin{equation}
\begin{aligned}
&p_{ZZ}(k_c,k_d|n_A,n_B) \\
=& \sum_{k=0}^{n_A} {n_A\choose k} \sum_{l=0}^{n_B} {n_B\choose l} {{\eta_A^k \eta_B^l (1-\eta_A)^{n_A-k}(1-\eta_B)^{n_B-l}}\over{2^{k+l}k!l!}} \\
&\sum_{m=0}^{k} {k \choose m}  \sum_{p=0}^{l} {l \choose p}\sum_{q=max(0,m+p-l)}^{min(k,m+p)} {k \choose q}{l \choose m+p-q}\\
&(m+p)!(k+l-m-p)!cos^{m+q}(\theta_A)cos^{m+p-q}(\theta_B)\\
&sin^{2k-m-q}(\theta_A)sin^{2l-m-2p+q}(\theta_B)\\
& - (1-\eta_A)^{n_A} (1-\eta_B)^{n_B} \\
\end{aligned}
\end{equation}

In the case with finite decoys (e.g. 3 decoy states for each of Alice and Bob), we can use linear programming to upper-bound the yields, which is described in more detail in Appendix A.
}

Afterwards, the phase-error rate can be upper-bounded using these yields:

\begin{equation}
\begin{aligned}
&p_{XX}(k_c,k_d)e_{ZZ}(k_c,k_d)\\
\leq& \sum_{i=0,1}\left[\sum_{n_A,n_B=0}^{\infty}c_{n_A}^{A,(i)}c_{n_B}^{B,(i)} \sqrt{p_{ZZ}(k_c,k_d|n_A,n_B)}\right]^2\\
\end{aligned}
\end{equation}

With the the X basis gain $p_{XX}(k_c,k_d)$, the X basis bit-error rate $e_{XX}(k_c,k_d)$, and the phase-error rate $e_{ZZ}(k_c,k_d)$, we can obtain the final secure key rate:

\begin{equation}
\begin{aligned}
R_{k_c k_d}&=p_{XX}(k_c,k_d)\\
&\times[1-h_2(e_{XX}(k_c,k_d))-h_2(e_{ZZ}(k_c,k_d))]\\
\end{aligned}
\end{equation}

\noindent where $h_2(x)=-xlog_2(x) - (1-x)log_2(1-x)$ is the binary entropy function.

\subsection{Effect of Channel and Intensity Asymmetry on Gain and QBER}

In the estimation of key rate, only three sets of observables are used: the X basis gain $p_{XX}(k_c,k_d)$, the X basis bit-error rate $e_{XX}(k_c,k_d)$, and the set of Z basis gain for each combination of decoy intensities $p_{ZZ}(k_c,k_d|\beta_A,\beta_B)$. Here we note that, the X basis gain and Z basis gain do not depend on channel asymmetry, and only the X basis QBER is affected by asymmetry.

For simplicity, here let us consider the second-order approximation for the Bessel function and exponential function, and for now ignore the phase mismatch and dark count rate: 

\begin{equation}
\begin{aligned}
I_0(x) &= 1 + {1\over 4}x^2 + O(x^4)\\
e^x &= 1 + x + {1\over 2}x^2 + O(x^3)\\
\end{aligned}
\end{equation}
\begin{figure*}[t]
	\begin{minipage}{\textwidth}
	\includegraphics[scale=0.24]{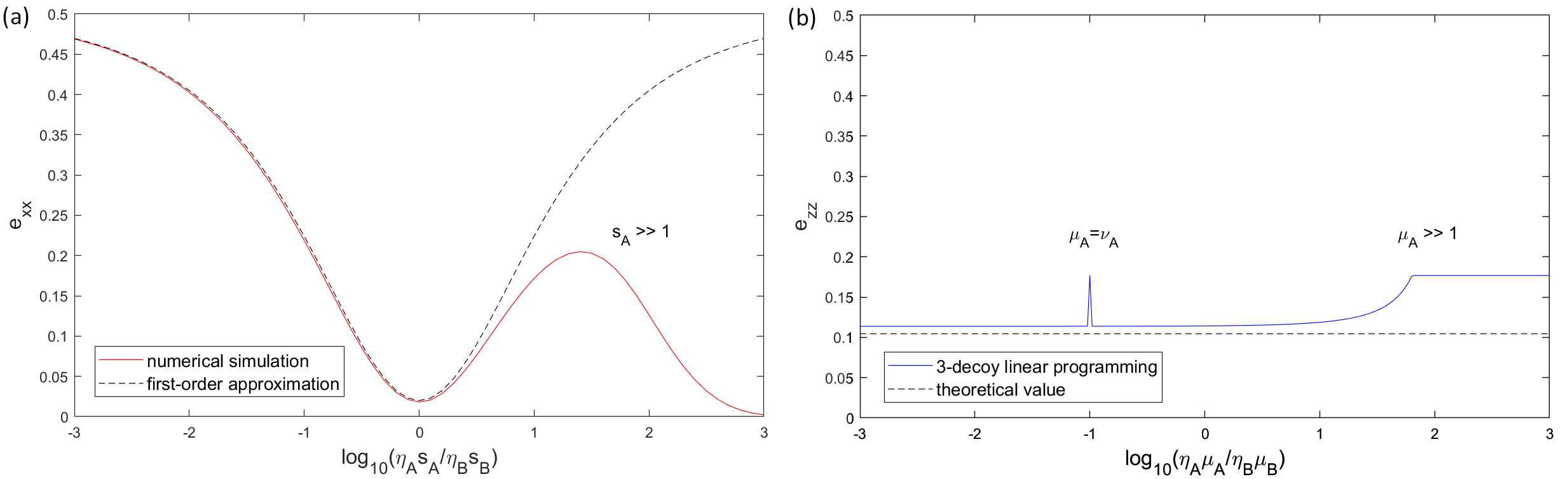}
	\caption{QBER versus asymmetry in arriving intensities at Charles, in the X and Z bases. Here we consider $k_c,k_d=0,1$, while the other case of $k_c,k_d=1,0$ has exactly the same values. Both plots use parameters $\eta_A=\eta_B=1$ (as well as detector efficiency $100\%$) and assume misalignment of $2\%$ from Alice and Bob each, and no dark counts or phase mismatch. (a) We plot the X basis QBER (acquired from full expression in Eq. (14)) as well as its first-order approximation (acquired from Eq. (24)). Here we vary $s_A$ while keeping $s_B=0.1$ to test different levels of asymmetry. As can be seen, the X basis QBER heavily depends on the symmetry between arriving intensities $s_A\eta_A$ and $s_B\eta_B$. Physically, this is because key generation depends on single-photon interference and therefore requires indistinguishability of incoming signals. When $\eta_B/\eta_A\neq 1$, X basis QBER will increase drastically if intensities are symmetric, while one can adjust $s_A/s_B$ such that $s_A/s_B=\eta_B/\eta_A$ to obtain minimal X basis QBER. \protect\footnotemark[1] (b) We plot the Z basis bit-error rate (i.e. X basis phase-error rate) obtained from linear programming using data from 3 decoy states. Here we set $\nu_A=\nu_B=0.01$, $\omega_A=\omega_B=0$, and signal states $s_A=s_B=0.1$. We fix $\mu_B=0.1$ and vary $s_A$ to test asymmetry. As can be seen, the upper-bounded phase error rate depends very little on the asymmetry between $\mu_A\eta_A$ and $\mu_B\eta_B$, and the linear program can effectively bound the error rate - in fact rather close to the theoretical value obtained with infinite-decoys - as long as $\mu_A,\mu_B$ are of reasonable values \protect\footnotemark[2]. {\color{black}The physical intuition is clear too: the yields are estimated by linear programming, which usually has redundant information (in the 9 cross terms $Q_{\mu_i\mu_j}$ where Alice and Bob each uses one of their three decoys) and is insensitive to asymmetry, and the phase-error rate (Eq. (17)) is a linear combination of the yields, which makes it insensitive to asymmetry just like the yields.}}
	\label{fig:QBER}
	\footnotetext{Note that first-order deviation no longer works for $s_A \gg 1$ - in this case, the majority of the detection events are double counts and are discarded, despite that the actual $e_{xx}$ is lower among the single detections (approaching zero as $s_A$ increases) - but generally the phase error rate estimation requires low signal intensities, and such high $s_A$ will not return positive key rate anyway, so this region will not be of interest to us.}
	\footnotetext{with two exceptions: when $\mu_A=\nu_A$, or $\mu_A \gg 1$, the constraints from observable data containing $\mu_A$ cannot provide any useful information, and the linear program has to use the data from one less decoy state, which is why the $e_{zz}$ is higher for these two extreme cases.}
\end{minipage}

\end{figure*}

We can then rewrite the X basis gain as:

\begin{equation}
\begin{aligned}
&p_{XX}(k_d,k_d)\\
=&{{1}\over{2}}\left( e^{-\sqrt{\gamma_A\gamma_B} cos\theta} + e^{\sqrt{\gamma_A\gamma_B} cos\phi cos\theta} \right) \times e^{-{{1}\over{2}}(\gamma_A+\gamma_B)}\\ & - e^{-(\gamma_A + \gamma_B)}\\
\approx& {1\over 2}(\gamma_A+\gamma_B) - {1 \over 8}\left[3 \gamma_A^2 + 3\gamma_B^2 + (2+4e_d)\gamma_A \gamma_B \right]\\
\end{aligned}
\end{equation}

\noindent and the Z basis gain as:

\begin{equation}
\begin{aligned}
&p_{ZZ}(k_d,k_d|\beta_A,\beta_B)\\
=&e^{-{{1}\over{2}}(\gamma'_A+\gamma'_B)}I_0(\sqrt{\gamma'_A\gamma'_B}cos\theta) - e^{-(\gamma'_A+\gamma'_B)}\\
\approx& {1\over 2}(\gamma'_A+\gamma'_B) - {1 \over 8}\left[3 \gamma_A^{'2} + 3\gamma_B^{'2} + (4+2e_d)\gamma'_A \gamma'_B \right]\\
\end{aligned}
\end{equation}

\noindent where the terms higher than second order are omitted, and $\theta$ is the total polarization misalignment angle between Alice and Bob satisfying $\theta = 2sin^{-1}(\sqrt{e_d}))$ (suppose Alice-Charles and Bob-Charles each has misalignment error $e_d$, but with misalignment angles in different directions). We can see that, the gain in both X and Z basis is dominated by the term ${1\over 2}(\gamma_A + \gamma_B)={1\over 2}(s_A\eta_A + s_B\eta_B)$ or ${1\over 2}(\gamma'_A + \gamma'_B)={1\over 2}(\mu_A^i\eta_A + \mu_B^j\eta_B)$, i.e. taking first-order approximation:

\begin{equation}
\begin{aligned}
&p_{XX}(k_d,k_d)\approx {1\over 2}(\gamma_A+\gamma_B)\\
&p_{ZZ}(k_d,k_d|\beta_A,\beta_B)\approx {1\over 2}(\gamma'_A+\gamma'_B)\\
\end{aligned}
\end{equation}

\noindent which means that the gain scales with the \textit{average} of arriving intensities through Alice's and Bob's channels - this is different from MDI-QKD, where the gain only contains the second-order terms $\gamma_A^2,\gamma_B^2,\gamma_A\gamma_B$. We can also see that the gain does not depend on the asymmetry of arriving intensities, e.g. $\gamma_A/\gamma_B$.

On the other hand, the QBER in X basis depends on the balance of arriving intensities:

\begin{equation}
\begin{aligned}
&e_{XX}(k_d,k_d)\\
=&{{ e^{-\sqrt{\gamma_A\gamma_B} cos\phi cos\theta}-(1-p_d)e^{-{{1}\over{2}}(\gamma_A+\gamma_B)}}  \over  {e^{-\sqrt{\gamma_A\gamma_B} cos\phi cos\theta} + e^{\sqrt{\gamma_A\gamma_B} cos\phi cos\theta}-2(1-p_d)e^{-{{1}\over{2}}(\gamma_A+\gamma_B)}}}\\
\approx& {{{1 \over 2}(\gamma_A+\gamma_B) - \sqrt{\gamma_A \gamma_B}cos\theta + {1\over 2}\gamma_A\gamma_Bcos^2\theta - {1\over 8}(\gamma_A + \gamma_B)^2}\over {(\gamma_A + \gamma_B) + \gamma_A\gamma_Bcos^2\theta - {1\over 4} (\gamma_A + \gamma_B)^2}}
\end{aligned}
\end{equation}

\noindent which, in the first-order approximation\footnote{The first order approximation for $e_{XX}(k_d,k_d)$ assumes that $\gamma_A,\gamma_B$ are much smaller than 1 - which is reasonable, since to get a good phase-error rate estimation, usually $s_A,s_B$ are smaller or equal to $0.1$, and for positions of interest where TF-QKD beats PLOB bound, the loss in each channel is usually larger than 10dB, which means that $\eta_A$ and $\eta_B$ are much smaller than 1 too - for instance 10dB channel loss corresponds to $0.1$ transmittance.}, can be simplified as:

\begin{equation}
\begin{aligned}
e_{XX}(k_d,k_d)\approx& {{{1 \over 2}(\gamma_A+\gamma_B) - \sqrt{\gamma_A \gamma_B}cos\theta}\over {\gamma_A + \gamma_B}}\\
=& {{{1 \over 2}({{\gamma_A \over \gamma_B}}+1) - \sqrt{{\gamma_A \over \gamma_B}}cos\theta}\over {{\gamma_A \over \gamma_B} + 1}}
\end{aligned}
\end{equation}

We can see that here the X basis QBER does depend on asymmetry - more precisely, it depends on how much the arriving intensities at Charles, $\gamma_A=\eta_As_A$ and  $\gamma_B=\eta_Bs_B$ are balanced. This is understandable physically, since the X basis key generation depends on single-photon interference and relies on the indistinguishability of incoming signals. This means that, in the case that channels are not symmetric, compensating for the channel asymmetry with different signal intensities for Alice and Bob and aiming for $\eta_As_A=\eta_Bs_B$  can help minimize the X basis QBER. 

On the other hand, in the Z basis, the bit-error rate (i.e. the X basis phase-error rate) cannot be directly measured, but is instead upper-bounded using the observable gain data from the decoy states. As we mentioned above, the Z basis gain (in the first-order approximation) scales with ${1\over 2}(\gamma'_A + \gamma'_B)={1\over 2}(\mu_A^i\eta_A + \mu_B^j\eta_B)$ and does not depend on the symmetry between incoming intensities. Moreover, the yields $p_{ZZ}(k_c,k_d|n_A,n_B)$ are estimated using linear programming. For instance, for three decoys where Alice and Bob respectively use $\{\mu_A,\nu_A,\omega_A\}$, $\{\mu_B,\nu_B,\omega_B\}$ as their decoy states, there are nine sets of observable gains, $\{Q_{\mu\mu},Q_{\mu\nu},Q_{\mu\omega},Q_{\nu\mu},Q_{\nu\nu},Q_{\nu\omega},Q_{\omega\mu},Q_{\omega\nu},Q_{\omega\omega}\}$, each of which constitutes a constraint for the linear program that helps bound the yields $p_{ZZ}(k_c,k_d|n_A,n_B)$. Such a structure makes the linear program relatively robust against asymmetry in the decoy states, and the linear program can fairly accurately upper-bound the yields as long as the intensities are of reasonable values (i.e. $\mu_A\neq \nu_A$, $\mu_B\neq \nu_B$, and none of the intensities are too large e.g. $>1$). 

{\color{black}
The phase error rate, as shown in Eq. (17), is based on a linear combination of the square root of the yields. It is therefore also very little affected by asymmetry, and almost always reaches a good value (at least in the infinite-data case) so long as the intensities are within reasonable range, regardless of the asymmetries in channel transmittances or decoy intensities.
}

We plot the QBER in the X and the Z bases versus asymmetry in arriving intensities (e.g. $s_A\eta_A/s_B\eta_B$ or $\mu_A\eta_A/\mu_B\eta_B$) in Fig. \ref{fig:QBER}. As can be seen, the X basis QBER depends heavily on asymmetry and is minimal when $s_A\eta_A/s_B\eta_B=1$, while the upper-bounded Z basis QBER (i.e. phase-error rate) is hardly affected by asymmetry.

Therefore, a viable strategy for TF-QKD in asymmetric channels is to compensate for the channel asymmetry with signal intensities $\{s_A,s_B\}$ only, while the decoy intensities $\{\mu_A,\nu_A,\omega_A\}$, $\{\mu_B,\nu_B,\omega_B\}$ can be still kept symmetric. However, note that the signal intensities not only determines (1) X basis QBER, it also affects (2) X basis gain (which determines the raw key generation rate, and favors large $s_A,s_B$), as well as (3) upper-bound of phase error rate (since the cat states are determined by signal intensities, and the estimation favors small $s_A,s_B$ - typically $<0.1$ - for a tighter upper bound on phase error rate). Criteria (1-3) cannot be simultaneously satisfied, therefore an optimization for $\{s_A,s_B\}$ is required for highest key rate.

Interestingly, we can compare this with the case of MDI-QKD. As described in Ref. \cite{mdi7intensity}, the 4-intensity protocol (and 7-intensity protocol in the extended asymmetric case) has decoupled X and Z bases, where Z basis is used for key generation and X basis uses decoy states to estimation phase-error rate. In MDI-QKD, the X basis data depends on two-photon interference and requires balanced arriving intensities (or else the X basis QBER will increase dramatically), while the Z basis does not require indistinguishability of the signals, and is therefore insensitive to channel asymmetry. In MDI-QKD, all the X basis decoy states should satisfy e.g. $\mu_A\eta_A = \mu_B\eta_B$, while the signal states $s_A,s_B$ can be chosen to simply optimize key generation rate. (Due to misalignment, there is a slight dependence of Z basis QBER to asymmetry too, hence optimal $s_A,s_B$ are still not equal, but this is a much weaker dependence on symmetry than in the X basis, and optimal $s_A/s_B$ is much closer to 1 than $\eta_B/\eta_A$ in MDI-QKD.) 

While our approach works both for MDI-QKD and TF-QKD, a key difference is that states that compensate for channel asymmetry are the signal states in TF-QKD (while this responsibility lies on decoy states in MDI-QKD), which are also involved in key generation and phase error estimation. This means that in TF-QKD, it is more difficult to simultaneously keep a low X basis QBER and a good key generation rate \& low phase error rate. Perhaps due to this reason, the advantage of asymmetric protocols is somewhat less pronounced in TF-QKD - nonetheless, it still provides about an order of magnitude higher key rate than completely symmetric protocols and still 2-3 times higher key rate than adding fibre - which means that it still is the strategy that provides highest key rate when channels are asymmetric.

\section{Numerical Results}
\begin{figure}[h]
	\includegraphics[scale=0.23]{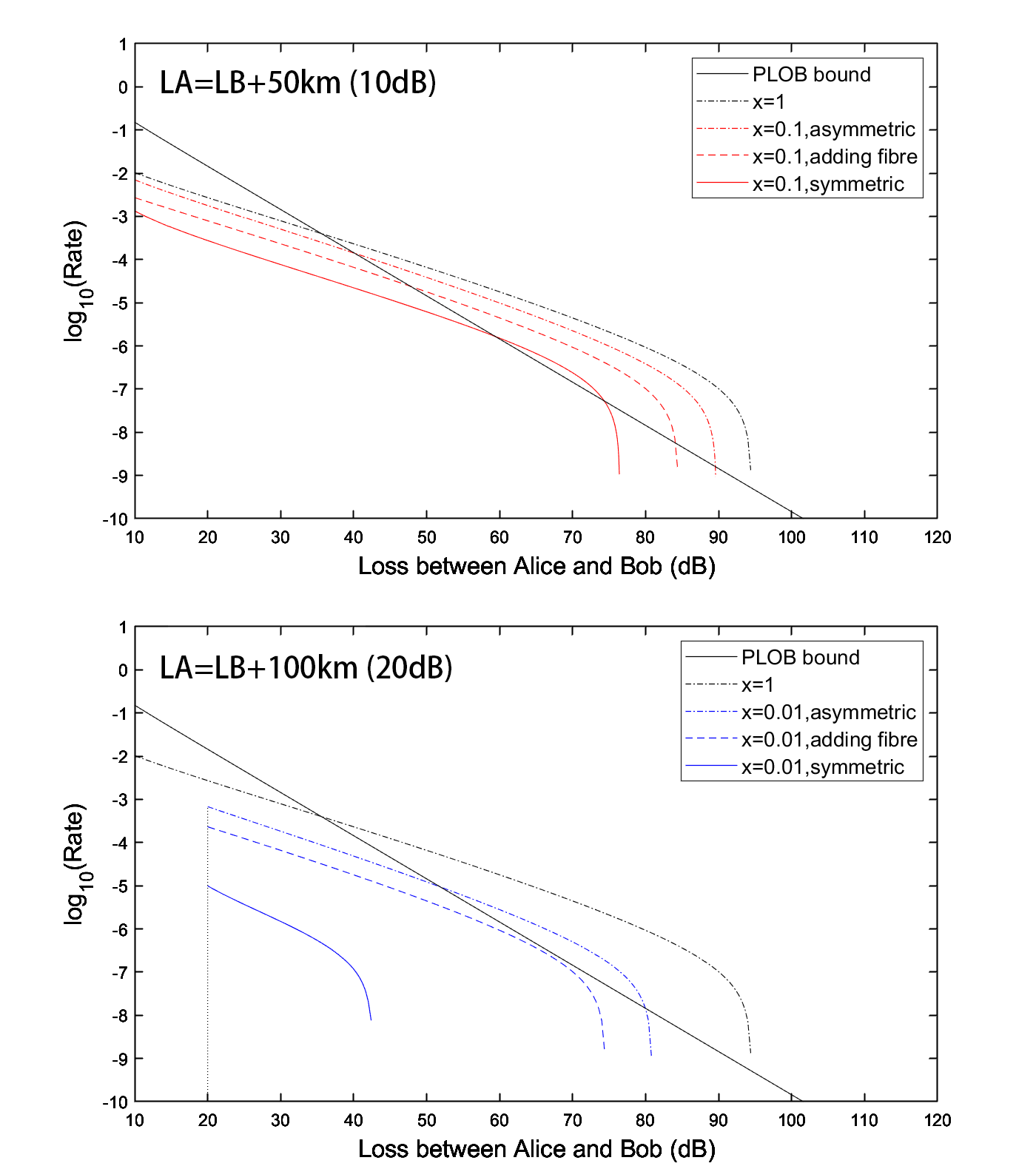}
	\caption{Key rate versus loss between Alice and Bob, for protocol with symmetric intensities ($s_A=s_B$), symmetric intensities with fibre added until channels are equal, and asymmetric intensities ($s_A,s_B$ fully optimized). The channel mismatch is fixed at $\eta_A/\eta_B=0.1$ (above) and $\eta_A/\eta_B=0.01$ (below), i.e. Alice-Charles always has 10dB (20dB) higher loss than Bob-Charles. The dark count rate is set to $10^{-8}$, and misalignment is $2\%$ for Alice and Bob each. As can be seen, allowing the use of asymmetric intensities greatly improves key rate when channels are asymmetric, and compared with a symmetric protocol, it can consistently provide approximately one order of magnitude higher key rate when there is a 10dB channel mismatch, and two orders of magnitude when there is a 20dB mismatch, for most distances. Interestingly, adding fibre can improve the key rate considerably too - but it still has lower key rate than the asymmetric protocol (the latter has about 2-3 times higher key rate), and has the additional inconvenience of having to modify the physical channel.}
	\label{fig:asym_rate}
\end{figure}

\begin{figure}[h]
	\includegraphics[scale=0.23]{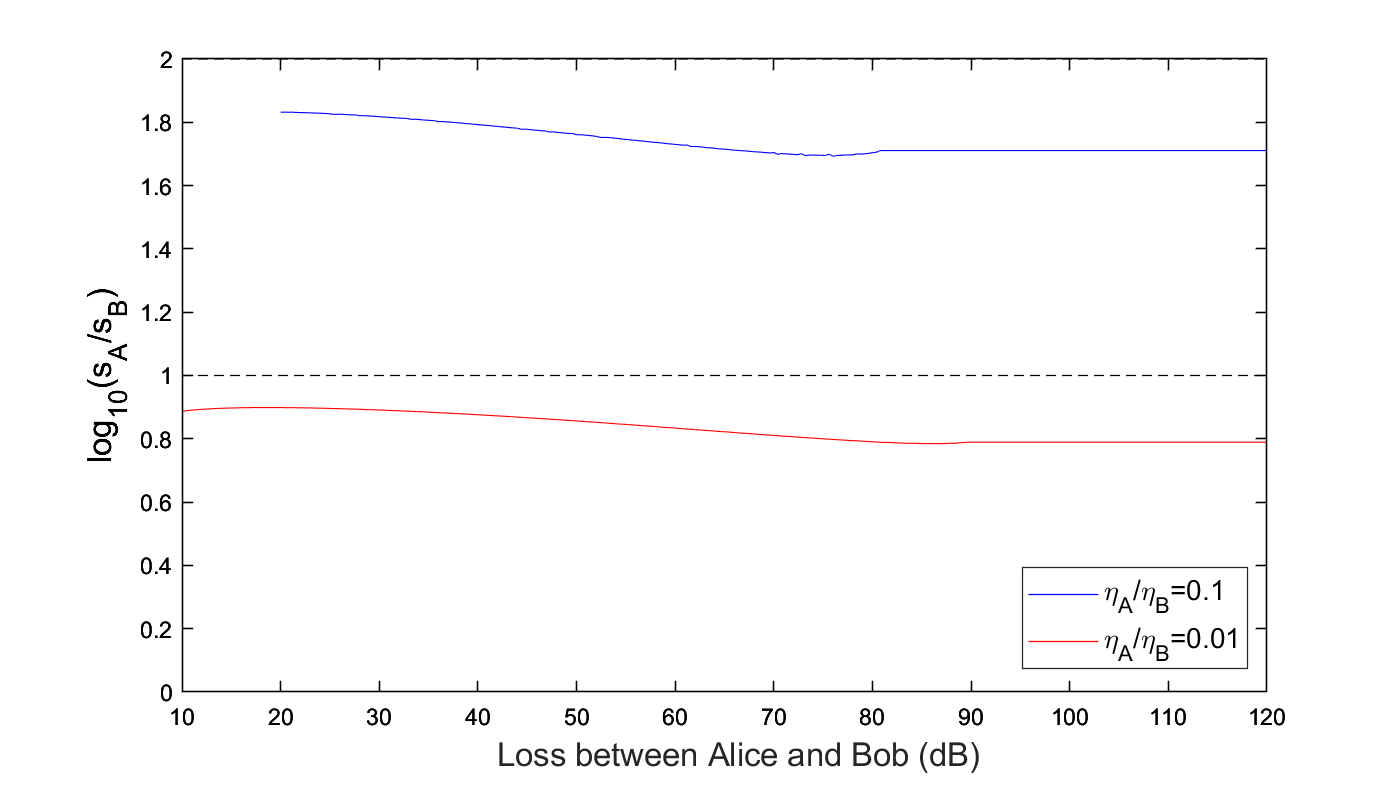}
	\caption{Ratio of optimal intensities $log_{10}(s_A/s_B)$ over loss between Alice and Bob. Here we test two cases where $\eta_A/\eta_B=0.1$ and $\eta_A/\eta_B=0.01$. As can be seen, $s_A/s_B$ is rather close to $\eta_B/\eta_A$ (here respectively $10^1$ and $10^2$). However, due to signal states being involved in key generation and phase-error rate estimation too, it slightly deviates from the value that minimizes X basis QBER (and instead takes the value that maximizes key rate).}
	\label{fig:compare_sA_sB}
\end{figure}

\begin{figure}[h]
	\includegraphics[scale=0.23]{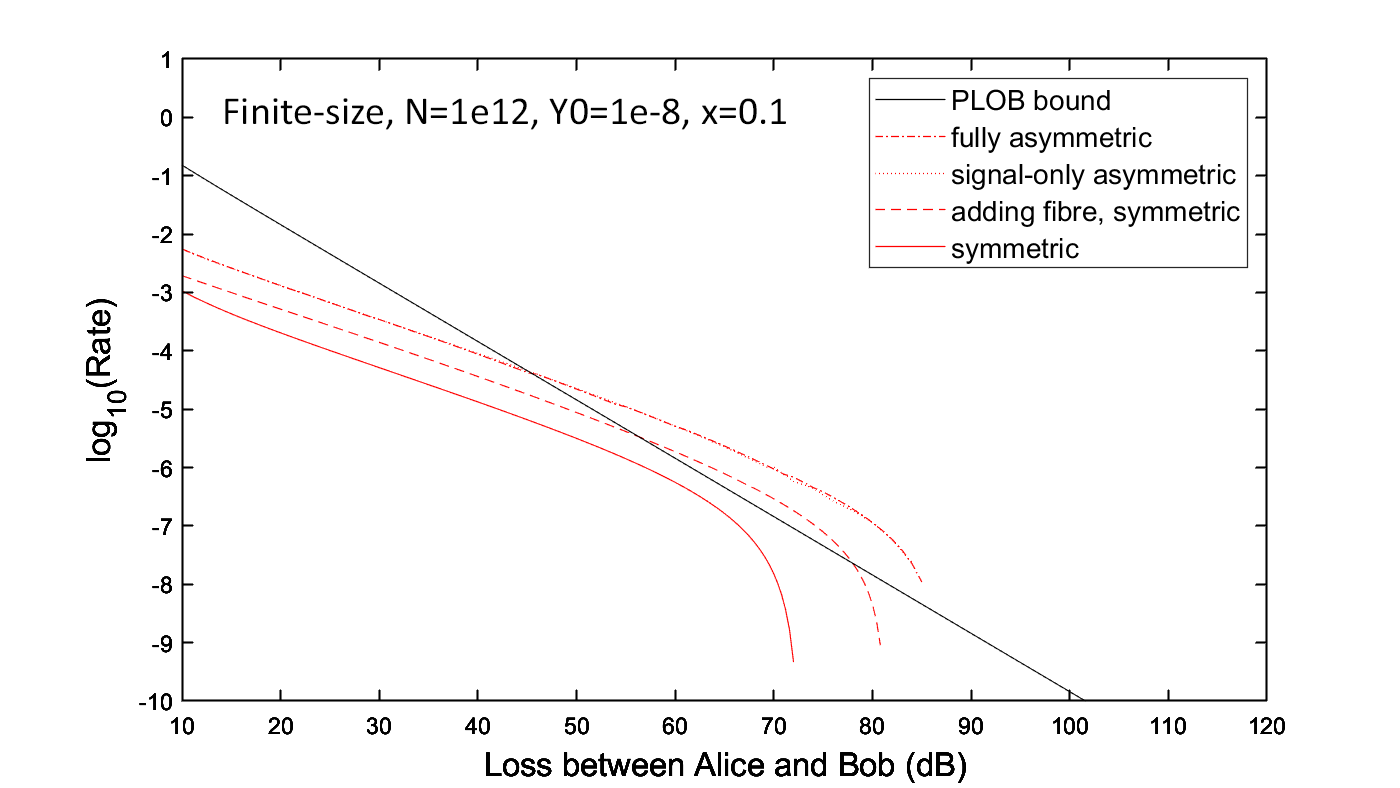}
	\caption{Key rate versus loss between Alice and Bob, for protocol with symmetric intensities and probabilities, with symmetric intensities and fibre added until channels are equal, with only asymmetric signal intensities (while all other parameters are symmetric between Alice and Bob), and with fully optimized parameters (all intensities and probabilities are freely optimized). Here we consider $10^{-8}$ dark count rate and $N=10^{12}$ total pulses sent, and channel asymmetry of $x=\eta_A/\eta_B=0.1$. As we can see, similar to the asymptotic case, using asymmetric intensities can greatly improve the key rate. Perhaps more interestingly, we can see that allowing asymmetry in signal intensities alone is sufficient in obtaining a good key rate through asymmetric channels (its key rate almost overlaps with the fully optimized case.)}
	\label{fig:rate_finite}
\end{figure}

In this section we use the technique described above - to compensate for channel asymmetry simply with different $s_A,s_B$ for Alice and Bob. We first compare our method with prior art techniques and study the numerically optimized intensities for the asymptotic (infinite-decoy, infinite-data) case. Then, we also show that our method works with finite decoys and also finite data size.

We plot the simulation results for asymptotic TF-QKD in Fig. \ref{fig:asym_rate}. As can be seen, for the two cases $\eta_A/\eta_B=0.1$ and $\eta_A/\eta_B=0.01$, our method consistently have much higher key rate than TF-QKD with symmetric intensities. Interestingly, we show that adding fibre can help users obtain higher key rate, but it comes with the additional inconvenience of having to physically modify the channel, and also it still has lower key rate than our method of simply adjusting signal intensities.

We also plot the ratio of optimal signal intensities in Fig. \ref{fig:compare_sA_sB}. As we have predicted, the optimal signal intensities are rather close to the relation of $s_A\eta_A=s_B\eta_B$, in order to maintain a lower X basis QBER. However, as we discussed, since signal states are also involved in key generation and phase error rate estimation (based on the imaginary cat states), they prevent the signal states from taking the values that minimize QBER (but rather, makes it choose the value that maximizes the overall key rate).

Additionally, we also plot our results for the practical case with finite number of decoys (here we use three decoys each for Alice and Bob: $\{\mu_A,\nu_A,\omega_A\}$, $\{\mu_B,\nu_B,\omega_B\}$) and finite data size. {\color{black} The upper-bounding of photon number yields using linear programming, as well as the finite-key analysis, are both described in more detail in Appendix A.} We can see that similar result holds - our method has an advantage over either using symmetric intensities directly or adding fibre. More interestingly, we include both the case where we only allow $s_A,s_B$ to be asymmetric, versus the case where all intensities and probabilities can be optimized, and as shown in the plot, we see that using asymmetric signal intensities alone is sufficient to compensate for channel asymmetry.

\section{Conclusion}

In this paper we present a simple method to obtain good performance for TF-QKD even if channels are asymmetric. We present a theoretical understanding of why signal states (and not decoy states) should be adjusted to compensate for asymmetry, and we also show that the method is still compatible with existing security proofs. With our method, there is no need to add additional fibre, and Alice and Bob can implement the method in software-only. This provides great convenience for TF-QKD in practice - where realistic channels might likely be asymmetric - and can also be used in quantum networks (where adding fibre for each pair of users is impractical) where a central service-provider can easily optimize the intensities for each pair of users.\\

\textit{Note added:} We note that, during the preparation of the current manuscript, it came into our knowledge that another work on asymmetric TF-QKD is under preparation \cite{tf02}, which is independently completed from this work.

\section{Acknowledgments}

This work was supported by the Natural Sciences and Engineering Research Council of Canada (NSERC), U.S. Office of Naval Research (ONR). Computations were performed on the Niagara supercomputer at the SciNet HPC Consortium. SciNet is funded by: the Canada Foundation for Innovation; the Government of Ontario; Ontario Research Fund - Research Excellence; and the University of Toronto.

{\color{black}

\appendix

\section{Numerically Estimating Photon-Number Yields with Linear Programs}

In this section we briefly describe the linear programming approach we used to estimate the upper bounds for the photon-number yields $p_{ZZ}(k_c,k_d|n_A,n_B)$ - which for simplicity here we will denote as $Y_{nm}$ - which is the probability to obtain a set of detection events $k_c,k_d$ given that Alice and Bob respectively sent $n_A,n_B$ (or, $n,m$) photons. Such an approach has been widely discussed in literature as in Refs.\cite{MDIAnalytical,mdiparameter,mdiChernoff}, and is also described in the simple TF-QKD proof paper \cite{simpleTFQKD}. We also used a similar linear programming approach for some of the results in Ref. \cite{mdi7intensity} Appendix E, but it was not described in detail in that paper.

For simplicity, in this section we denote the observable gain in Z basis $p_{ZZ}(k_c,k_d|\beta_A,\beta_B)$ as $Q_{\mu_i,\mu_j}$ where $\mu_i=\beta_A^2$ and $\mu_j=\beta_B^2$, and $k_c,k_d$ are omitted (since the same expressions hold true for $k_c,k_d=(0,1)$ or $k_c,k_d=(1,0)$, and we can substitute the observable data for each $k_c,k_d$ respectively to obtain the corresponding $p_{ZZ}(k_c,k_d|n_A,n_B)$). Also, as mentioned above, we denote the yields $p_{ZZ}(k_c,k_d|n_A,n_B)$ as $Y_{nm}$.

\subsection{Linear Program Model}

Following Ref.\cite{mdiqkd}, the yields $Y_{nm}$ where Alice sends n photons and Bob sends m photons, satisfy the constraints:

\begin{equation}
\begin{aligned}
\sum_{n}\sum_{m}P^{\mu_i}_{n}P^{\mu_j}_{m}Y_{nm} &= Q_{\mu_i\mu_j}^Z \\
\end{aligned}
\end{equation}

\noindent where the photon number distributions are Poissonian:

\begin{equation}
\begin{aligned}
P^{\mu_i}_{n} &= e^{\mu_i} {{\mu_i^n}\over {n!}}\\
P^{\mu_j}_{m} &= e^{\mu_j} {{\mu_j^m}\over {m!}}\\
\end{aligned}
\end{equation}

Here, the right-hand-side constants $Q_{\mu_i\mu_j}^Z$ are the "observables", i.e. the gain and error-gain respectively for the intensity combination $\mu_i,\mu_j$ (which can be any intensity among the set of decoy intensities). For the case of 3-decoys each for Alice and Bob, Eq. (A1) corresponds to 9 sets of constraints. Using Eq. (A1) as linear constraints, and $\{Y_{nm}\}$ as variables, we can apply linear programming, to maximize or minimize any linear combination of any of the variables (called an objective function) - for instance, here we can run the linear program multiple times, each time acquiring the upper bound for a given $Y_{nm}$ where $(n,m)$ can be $(0,0),(2,0),(0,2),(1,1),(2,2)$.

Note that, since there are infinitely many photon number states, to solve the linear program on an actual computer, we have to perform a cut-off and discard higher-order terms with large photon number. In practice we choose $S_{cut}=10$, such that a term is only discarded when both $n\geq 10$ and $m\geq 10$. For the discarded terms, we can either set them to zero (for lower bounds) or 1 (for upper bounds). 

\begin{equation}
\begin{aligned}
\sum_{n}\sum_{m}P^{\mu_i}_{n}P^{\mu_j}_{m}Y_{nm} &\geq \sum_{n < 10}\sum_{m < 10}P^{\mu_i}_{n}P^{\mu_j}_{m}Y_{nm} \\
\sum_{n}\sum_{m}P^{\mu_i}_{n}P^{\mu_j}_{m}Y_{nm} &\leq \sum_{n < 10}\sum_{m < 10}P^{\mu_i}_{n}P^{\mu_j}_{m}Y_{nm}\\
&+ \left(1-\sum_{n < 10}\sum_{m < 10}P^{\mu_i}_{n}P^{\mu_j}_{m}\right)\\
\end{aligned}
\end{equation}

Therefore, in practice, the linear constraints can be written as:

\begin{equation}
\begin{aligned}
&Q_{\mu_i\mu_j}^Z - \left(1-\sum_{n < 10}\sum_{m < 10}P^{\mu_i}_{n}P^{\mu_j}_{m}\right) \\
&\leq \sum_{n < 10}\sum_{m < 10}P^{\mu_i}_{n}P^{\mu_j}_{m}Y_{nm} \\
&\leq Q_{\mu_i\mu_j}^Z\\
\end{aligned}
\end{equation}

\noindent with the additional constraint on variables:

\begin{equation}
\begin{aligned}
 0 \leq Y_{nm} \leq 1 \\
\end{aligned}
\end{equation}

The linear program is run multiple times, each time maximizing a given $Y_{mn}$, where $(n,m)$ can be $(0,0),(2,0),(0,2),(1,1),(2,2)$.

\subsection{Finite-Size Effects}

In this subsection we consider finite-size effects for the privacy amplification process. Because of the statistical fluctuations, the observables (gains) we obtain in the Z basis might deviate from their respective expected values, which will lie within a certain ``confidence interval" around the observed values. Here we will perform a standard error analysis, similar to that in \cite{mdiparameter,mdifourintensity,mdi7intensity}, which is meant to be a straightforward estimation of the performance of TF-QKD under asymmetry and with practical data size, but not as a rigorous proof for composable security. 

Consider a random variable, whose observed value is $n$, we can bound its expected value $\langle n \rangle$ with the upper and lower bounds

\begin{equation}
\begin{aligned}
\underline{n} = n - \gamma \sqrt{n} \leq \langle n \rangle \leq  n + \gamma \sqrt{n} = \overline{n}
\end{aligned}
\end{equation} 

\noindent with a confidence (success probability) of $erf({\gamma/\sqrt{2}})$, where $\gamma$ is the number of standard deviations the confidence interval lies above and below the observed value, and $erf$ is the error function. In the simulations we consider a security failure probability of $\epsilon=10^{-7}$, which means we should set $\gamma \approx 5.3$.

In the Z basis, let us denote the observed counts for a given intensity setting $\{\mu_i,\mu_j\}$ as $n_{\mu_i,\mu_j}^Z$, which satisfies

\begin{equation}
\begin{aligned}
n_{\mu_i,\mu_j}^Z = Q_{\mu_i,\mu_j}^Z \times (NP_{\mu_i}P_{\mu_j})
\end{aligned}
\end{equation}

\noindent where $N$ is the total number of signals sent and $P_{\mu_i},P_{\mu_j}$ are the probabilities for Alice and Bob to respectively choose intensities $\mu_i$ and $\mu_j$. By applying Eq. (A6), we can acquire the upper and lower bounds to $Q_{\mu_i,\mu_j}^Z$:

\begin{equation}
\begin{aligned}
\overline{Q_{\mu_i\mu_j}^Z}=Q_{\mu_i\mu_j}^Z + \gamma \sqrt{Q_{\mu_i\mu_j}^Z \over {N P_{\mu_i} P_{\mu_j}}} \\
\underline{Q_{\mu_i\mu_j}^Z}=Q_{\mu_i\mu_j}^Z - \gamma \sqrt{Q_{\mu_i\mu_j}^Z \over {N P_{\mu_i} P_{\mu_j}}} \\
\end{aligned}
\end{equation}

Then, we can substitute them into the upper and lower bounds in the linear program when estimating $Y_{nm}$:

\begin{equation}
\begin{aligned}
&\underline{Q_{\mu_i\mu_j}^Z} - \left(1-\sum_{n < 10}\sum_{m < 10}P^{\mu_i}_{n}P^{\mu_j}_{m}\right) \\
&\leq \sum_{n < 10}\sum_{m < 10}P^{\mu_i}_{n}P^{\mu_j}_{m}Y_{nm} \\
&\leq  \overline{Q_{\mu_i\mu_j}^Z}\\
\end{aligned}
\end{equation}

\noindent which loosens the bounds and will result in a slightly higher upper bound for $Y_{nm}$ (which is understandable, since we expect lower key rate with finite-size effect considered). Similar linear programs for finite-size decoy-state have also been considered in Ref. \cite{MDIAnalytical}.

Note that, although here we only consider a standard error analysis, in principle our results in this paper is applicable to e.g. composable security using Chernoff's bound \cite{mdiChernoff}. The key point is, the dependence on channel asymmetry, and the compensation for asymmetry using intensities, are only relevant in the X basis (signal states). The asymptotic case (with infinite decoys, where only signal states are relevant) therefore defines the fundamental scaling of key rate versus asymmetric channels, and all types of finite size analysis on the decoy states (e.g. using standard error analysis, using Chernoff's bound \cite{mdiChernoff}, adding a joint bound analysis to tighten the bounds\cite{mdifourintensity}, or not using finite-size analysis at all) can be viewed of as correction terms (imperfections) on the yields and the key rate. Our method is only related to the signal states and their intensities in the X basis, and is in principle always applicable regardless of the type of decoy state analysis (e.g. number of decoys) and the finite-size analysis used, as long as the Z basis is decoupled from the X basis.

With finite-size effect considered, the optimizable parameters for TF-QKD now include 

\begin{equation}
\begin{aligned}
	[&s_A,\mu_A,\nu_A,P_{s_A},P_{\mu_A},P_{\nu_A},\\
&s_B,\mu_B,\nu_B,P_{s_B},P_{\mu_B},P_{\nu_B},]
\end{aligned}
\end{equation}

\noindent where the implicit parameters are $\omega_A,\omega_B$ (which for simplicity we assume to be zero), and $P_{\omega_A}=1-P_{s_A}-P_{\mu_A}-P_{\nu_A}$ and similarly $P_{\omega_B}=1-P_{s_B}-P_{\mu_B}-P_{\nu_B}$, and the choice of signal states $s_A,s_B$ versus the decoy states automatically implies basis choice, too. The above parameters are optimized using the same coordinate descent algorithm as described in Ref. \cite{mdi7intensity}. In Fig. \ref{fig:rate_finite}, the dot-dash line (fully asymmetric) optimizes all 12 parameters, while the dashed line (signal-only asymmetric) optimizes only 7 parameters (where all parameters except $s_A,s_B$ are identical for Alice and Bob):

\begin{equation}
\begin{aligned}
[&s_A,\mu,\nu,P_{s},P_{\mu},P_{\nu}, s_B,\mu,\nu,P_{s},P_{\mu},P_{\nu}]
\end{aligned}
\end{equation}

Performing coordinate descent on key rate versus parameters while estimating the yields with linear programming is rather CPU-intensive. We have used a 40-core (80-thread) machine (a single compute node in the Niagara supercomputer \cite{SciNet}, each node with dual 20-core Intel Skylake CPUs) to generate Fig. \ref{fig:rate_finite}, where the OpenMP multithreading library is used to parallelize the coordinate descent algorithm (to accelerate the search along each coordinate). The details of the algorithm can be found in Ref. \cite{mdi7intensity,mdiparameter}. Also, we used Gurobi \cite{gurobi}, a commercial linear program solver, to solve the linear programming models. Linear programs sometimes introduce multiple maxima, which means a local search on parameters sometimes might get trapped in a local maximum. To alleviate this, we can start a local search from multiple random starting points, and pick the largest search result, which can be viewed of as a form of global search. (In principle, we can permutate the search results and perform multiple iterations of random search using e.g. an evolution algorithm \cite{evolution}, but here using one iteration with multiple random starting points is usually sufficient in finding a good key rate).

}
\end{document}